\documentclass[twocolumn]{aastex631}
\submitjournal{JAAVSO}

\shorttitle{Characterizing Photonic Ring Resonators}
\shortauthors{Hermann \textit{et al.}}
\graphicspath{{./}{figures/}}
\usepackage{amsmath}  % needed for \tfrac, \bmatrix, etc.
\usepackage{amsfonts} % needed for bold Greek, Fraktur, and blackboard bold
\usepackage{graphicx} % needed for figures
\usepackage{makecell}
\usepackage{xcolor}

\usepackage{color,soul}
\usepackage{natbib}
\newcommand{\SNIa}{SN\,Ia }
\newcommand{\SNeIa}{SNe\,Ia}

\begin{document}

\begin{singlespace} 

% Be sure to use the \title, \author, \affiliation, and \abstract macros
% to format your title page.  Don't use lower-level macros to  manually
% adjust the fonts and centering.

\title{Characterizing Photonic Ring Resonator Filters for OH Suppressed Near-infrared Astronomy}

%When submitting the manuscript for review, do not include the author's name or institution
\author{A. Hermann}
\email{amynhermann@gmail.com} % optional
\author{J. Lasker}
\author{R. Vaisakh}
\author{R. Kehoe}
\author{R. Staten}
\author{A. Wallace}
\affiliation{Department of Physics, Southern Methodist University, Dallas, Texas 75205}

\author{S. Ellis}
\affiliation{Australian Astronomical Observatory, North Ryde, New South Wales 2113, Australia}

\author{K. Kuehn}

\affiliation{Lowell Observatory, Flagstaff, Arizona 86001 }

\author{S. Kuhlmann}
\affiliation{Argonne National Laboratory, Lemont, Illinois 60439}

\date{\today}

%When submitting the manuscript for review, do not include the author's name or institution

\date{\today}

%%%%%%%%%%%%%%%%%%%%%%%%%%%%%%%%%%%%%%%%%%%%%%%%
\begin{abstract}
Supernova cosmology relies on accurate measurement of the absolute magnitudes of Type Ia supernovae. Observation in the visible incurs significant systematic uncertainties in these measurements due to their high degree of interstellar dust extinction. Observing in the near-infrared (750-2500 nm) mitigates this issue by decreasing attenuation, decreasing dispersion, and increasing supernova observation distances. However, ground-based observations in the near-infrared suffer from sky background caused by atmospheric OH emission lines. Advancements in photonic ring resonator filter devices make it feasible to suppress these lines. In this paper, we evaluate the performance of a prototype photonic ring resonator device developed for use in the lower H-band (1480-1620 nm). Specifically, we characterize the progression of suppressed wavelengths into the upper H-band (1620-1800 nm) enabling the suppression of emission lines over a broader wavelength range.  
\end{abstract}

%\maketitle % title page is now complete

%%%%%%%%%%%%%%%%%%%%%%%%%%%%%%%%%%%%%%%%%%%%%
\section{Introduction}
\label{sec:intro}

Astronomical sources are detectable when a source's incident flux is higher than its background flux. Currently, long exposure sky-subtraction techniques can be employed to detect faint sources \citep{morganson_2018,blanton_2011}. One significant source of background flux in the near infrared (NIR, 750-2500 nm) arises from molecular emission in the atmosphere, called sky background. In the H-band ($\lambda_{\rm eff}$, the throughput-weighted average wavelength of a filter throughput function, is $\sim$1650 nm for the H-band), re-emission of light through atmospheric OH radicals creates an abundance of emission lines that impede NIR measurements. \citet{rousselot_2000} provides a computed spectrum demonstrating OH emission's impacts on the observed H-band. These OH emission lines increase photon noise in observed spectra and, accordingly, decrease measurements' signal-to-noise ratio (SNR). As a result, ground-based observations in the NIR require impractically long exposure times that would harm measurements' sensitivity to time domain signals. Additionally, standard sky-subtraction techniques are challenging in the NIR because their backgrounds experience both long and short temporal variability. Long term temporal variability arises from OH molecules only exciting during the day. Excited OH molecules slowly de-excite overnight, causing the background emission spectra to gradually weaken. Conversely short term temporal variability, typically on the scale of minutes, results from random atmospheric pressure changes. Both types of variability in emission spectra make sky subtraction methods difficult to implement \citep{scellis_2008}. \par

This complication is problematic for the measurement of dark energy with Type Ia supernovae (\SNeIa{}), which depends on accurate measurement of their absolute magnitudes for use as standardizable candles \citep{Avelino19,barone-nugent_2012}. Observing \SNeIa{} in the visible leads to large systematic uncertainties due to variability from interstellar dust extinction \citep{brout_2022}. NIR observation mitigates this issue by minimizing the impacts of attenuation and scatter from interstellar dust. As demonstrated by \citet{Wood-Vasey08}, \citet{Phillips12},  \citet{Avelino19}, and \citet{Peterson24} the RMS scatter of \SNeIa{} distances measured from photometric surveys is significantly reduced when light curve fits incorporate information from the Y($\lambda_{\rm eff} \sim1000$ nm), J($\lambda_{\rm eff} \sim1250$ nm), H ($\lambda_{\rm eff} \sim1650$ nm), and K ($\lambda_{\rm eff} \sim2200$ nm) bands, spanning the NIR. The most recent of these studies  \citep{Peterson24} showed a reduction in \SNIa{} Hubble Diagram distance modulus residual RSD (robust median absolute standard deviation) by up to 0.047 mag when considering YJH band data as compared to only considering visible (Sloan $gri$ band or ATLAS $co$ band) data for the same set of \SNeIa{}, with the H to $gri$ band comparison showing the largest difference. \citet{Peterson24} also finds a small reduction in the size of the so-called ``mass-step," a step discontinuity in estimated \SNIa luminosities with respect to their host masses, with the H band once again showing the smallest and least statistically significant mass step. These benefits of observing in the NIR would improve systematic uncertainties of The Vera Rubin Observatory Legacy Survey of Space and Time (LSST). While LSST does not fully extend into the infrared, this paper seeks to pursue research and development of ring resonator filters for use in such a follow up instrument. To best utilize the NIR's advantages, the sky background impacts of atmospheric OH emission lines must be mitigated. Space-based observations provide one possible solution for the elimination of sky background from OH emission lines. However, this is not a widely accessible solution due to cost and design challenges.\par

Since each emission line's wavelength is known, it is possible to suppress these emission lines in a measured spectrum --- a method termed sky background suppression. This method lowers the sky background flux, achieving ground-based observations with better SNRs. However, background suppression requires high precision; even a misalignment of a fraction of a nanometer can impede the suppression of emission lines. PRR filters for OH suppression have dips with full widths at half max (FWHM) of 0.2 nm for adequate OH suppression \citep{pliu_2021,kkuehn_2020}. High precision astronomical spectrographs require 4-6 samples per FWHM for accurate reconstruction of a symmetric line shape \citep{Robertson_2017, Chance_2005}. We employ ten samples across the linewidth such that there is sufficient information about each dip's shape to ensure total suppression regardless of symmetry. Accordingly, a resolution of 20 pm is used to ensure complete suppression. \par

Two filtering methods are currently under investigation for background suppression: photonic ring resonators (PRRs) and fiber Bragg gratings (FBGs). FBGs utilize refractive index modulation to reflect back narrow bands of a specified wavelength, while allowing the remaining light to transmit through the fiber \citep{FBG_text}. However, the aperiodic FBGs needed for OH suppression are expensive and do not have the ability to tune for primary sources of maladjustment, e.g. by using athermal tuning \citep{yoffe_1995}. In the event of small manufacturing errors, the lack of tunability would prevent accurate placement of dips that could harm the quality of background suppression. PRRs, conversely, eliminate specific wavelengths (called suppressed wavelengths) through their circular waveguides as described in Section \ref{sec:theory}. Combining several PRRs in series suppresses multiple wavelengths \citep{microres_text}. In PRRs the suppressed wavelength is set during fabrication, but can be adjusted after fabrication by heating or cooling the PRR. Additionally, PRRs are modular, easily scaled, and can be as much as ten times smaller than FBGs. As a result, series of PRRs are more easily implemented in compact packages than FBGs.\par
For these reasons, we studied PRRs' utility in suppressing NIR sky background. Previous work in the field has focused on measuring the performance of PRR devices in the lower H-band (1480-1620 nm) \citep{pliu_2021,kkuehn_2020,kkuehn_2024}. In this paper, we expand the study of PRR performance in background suppression into the upper H-band (1620-1800 nm). To achieve this, we construct an experimental setup to determine the performance of a PRR device for a subsection of the upper H-band (1620-1665 nm) and then characterize its performance. With the collected data, we extrapolate on a constructed fit to quantify the evolution of different performance metrics, particularly the suppressed wavelengths, from the lower H-band into the upper H-band.

\section{PRR Theory \& Fabrication}
\label{sec:theory}

\begin{figure}[h!]
\centering
\vspace{3pt}
\includegraphics[scale=0.4]{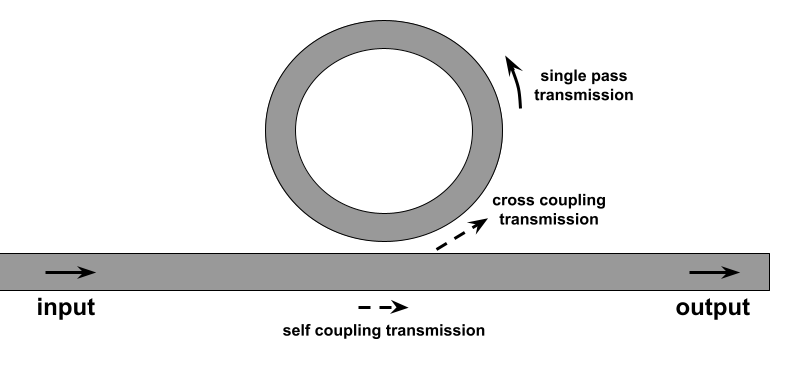}
\caption{Diagram provides an abstract representation of the evanescent coupling between the ring resonator and the source fiber. Solid arrows indicate signals, while dashed arrows indicate coupling coefficients.}
\label{ring_resonator}
\end{figure}

A photonic ring resonator consists of a circular waveguide and a source fiber that are evanescently coupled to suppress individual wavelengths. Evanescent coupling allows some energy -- in the form of an evanescent wave -- to transmit with an exponential decay between closely spaced waveguides, particularly in conditions at or near total internal reflection (TIR) \citep{hecht}.  Figure \ref{ring_resonator} showcases the basic structure of a PRR. \par

As a result of evanescent coupling, the entire spectrum of input light cross-couples into the ring. Within the ring, each wavelength undergoes repeated TIR. For the case of negligible attenuation from coupling, wavelengths that do not meet the resonance condition (Eq. \ref{resonance_condition}) are destructively interfered,

\begin{equation}
\label{resonance_condition}
\phi = 2\pi m
\end{equation}

\noindent where $\phi$ is the input signal's phase at a given wavelength and $m$ is a positive integer called the order. The wavelength at which the resonance condition is met for $m=1$ is termed the first order resonant wavelength ($\lambda_{res}$). Subsequent resonance conditions occur at integer multiples of the ring's circumference. These higher order resonant wavelengths ($\lambda_{res,m}$) couple into the loop, becoming the single-pass transmission shown in Fig. \ref{ring_resonator}. Here, they form a series of standing waves created from the original injected signal \citep{bogaerts_2011}. Upon re-coupling with the waveguide, the $\pi$ phase shift from the single pass transmission in the loop induces destructive interference at the resonant wavelengths in the input signal (See Eq. 3.15 in \cite{microres_text}). Thus, the PRR's output signal contains dips at each resonant wavelength in the output spectrum; the wavelengths at which those dips occur are also called suppressed wavelengths ($\lambda_{sup,m}$). The spacing between consecutive modes of resonant frequencies is referred to as the free spectral range ($FSR_\lambda$). For a single ring resonator, the resonant wavelength of the mode and the free spectral range are determined by the PRR's circumference ($L$), the group refractive index ($n_g$), the order ($m$), and the free space wavelength ($\lambda_0$) of the ring resonator \citep{microres_text}.

\begin{equation}
\label{resonant_wavelength}
\lambda_{res,m} =m\lambda_{res}= mLn_g
\end{equation}

\begin{equation}
\label{free_spectral_range}
FSR_\lambda = \frac{\lambda_0^2}{Ln_g}
\end{equation}

The order is necessarily a positive integer. The group refractive index ($n_g$) quantifies the phase delay in the waveguide per unit of length; it is dependent upon the material of transmission \citep{microres_text}. \par

\section{Apparatus \& Data}

A single PRR, however, can only suppress one OH emission line at a time. The simultaneous suppression of multiple emission lines requires the coupling of multiple PRRs. Previous work by our collaboration produced a device of five in-series PRRs such that five lines in the lower H-band could be suppressed. These five PRRs were embedded within a silicon wafer that was fiber-coupled with astronomical instrumentation \citep{pliu_2021,kkuehn_2020}. This silicon wafer was designed at Argonne National Lab, fabricated by Applied Nanotools, Inc, and termed ANT-07.\par

\begin{figure}[h!]
\centering
\vspace{3pt}
\includegraphics[scale=0.35]{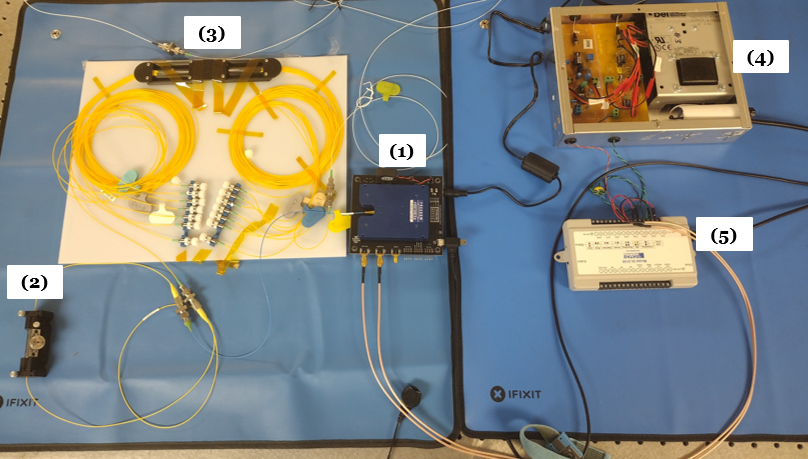}
\caption{(1) tunable laser with wavelength range of 1620 to 1665 nm, (2) adjustable polarization controller, (3) photonic ring resonator device ANT-07, (4) InGaAs detector, (5) digitizer. All components are connected via fiber coupling.}
\label{experimental_setup}
\end{figure}

 The collaboration then constructed a setup to test ANT-07 in the lower H-band. We constructed a similar experimental setup to test ANT-07 in the upper H-band (see Fig. \ref{experimental_setup}). In both setups, a tunable laser was fiber-optically coupled to an in-line fiber optic polarization controller and, later, ANT-07. The output from ANT-07 was recorded by an InGaAs detector (designed by Dave Underwood at Argonne National Lab) which was subsequently fed into a digitizer. In the lower H-band setup, the tunable laser was a TLK-1550R from ThorLabs with a range of 1480 to 1620 nm. In the upper H-band, we used the FP4209 InstaTune Fast Random-Access Tunable Laser from Freedom Photonics with a range of 1620 to 1665 nm. \par

\begin{figure}[h!]
\centering
\vspace{3pt}
\includegraphics[scale=0.25]{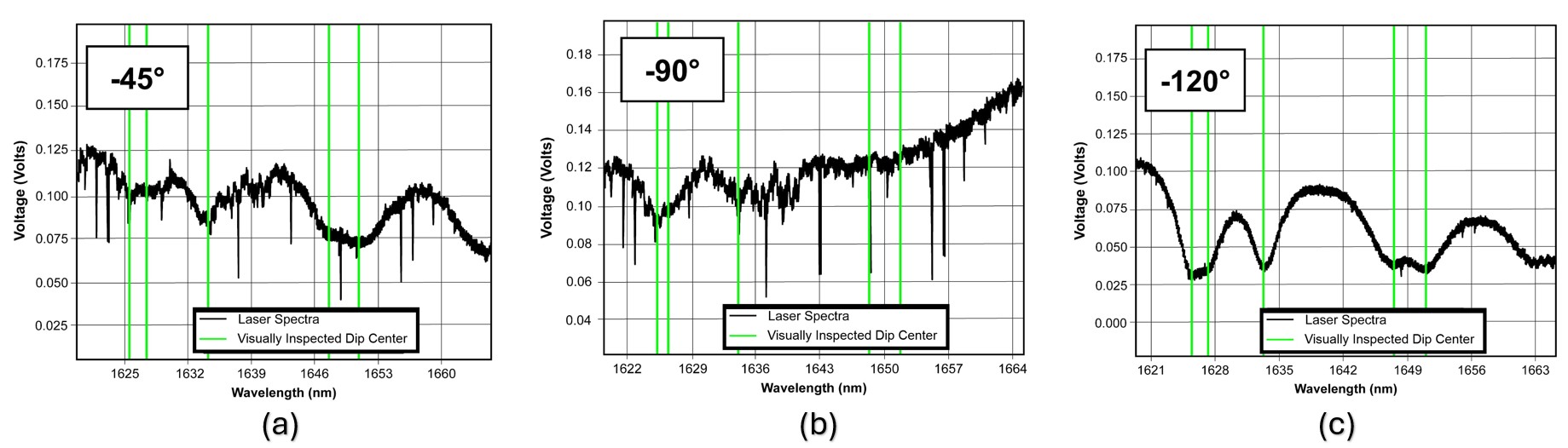}
\caption{Unflattened spectra of ANT-07 in the upper H-band. All dips are in their sixth order. The relative polarization angles are (a) -45° (b) -90° (c) -120°. Spectrum (c) was selected for subsequent upper H-band analysis. Green lines indicate the visually inspected dip locations from (c), as later determined in Table \ref{wavelength_preds_exp} and shown in Fig. \ref{upperH_spectra}. Spectra (a) and (b) have these dip locations overlaid, demonstrating how polarization artifacts can obscure the true dip locations. Please note that angles are approximate due to setup limitations.}
\label{upperH_polarized}
\end{figure}

Additionally, the upper H-band experimental setup included a protractor to measure the angle of the in-line polarization controller, which we varied incrementally to determine the angle at which the dips were strongest. We collected data using this setup for several relative polarization angles, including -120°, -90°,-45°, 0°, 45°, and 90°. These increments are selected due to the coarse precision of our measurements. It is crucial to note that the angle of the in-line polarizer does not directly describe the polarization of the transmitted signal. Rather, it allows for the selection of specific angles that maximize the output spectrum's SNR. Fig. \ref{upperH_polarized} contains some representative measurements.

Based off these spectra and an analysis of their SNRs, the in-line polarizer angle was set to -120° for all subsequent upper H-band measurements. This selection minimized the polarization related effects on spectra, which can obscure true dip locations (as seen in Figs. \ref{upperH_polarized}a and \ref{upperH_polarized}b). Additionally, the dips are quite broad at the sixth order of suppression (Fig. \ref{upperH_polarized}c). Section \ref{sec:Analysis} will discuss the causes and consequences of this broadening with reference to the lower H-band spectra (Fig. \ref{lowerH_spectra}) and the upper H-band spectra (Fig. \ref{upperH_spectra}). 

\section{Analysis \& Results}
\label{sec:Analysis}
\begin{figure}[h!]
\centering
\vspace{3pt}
\includegraphics[scale=0.23]{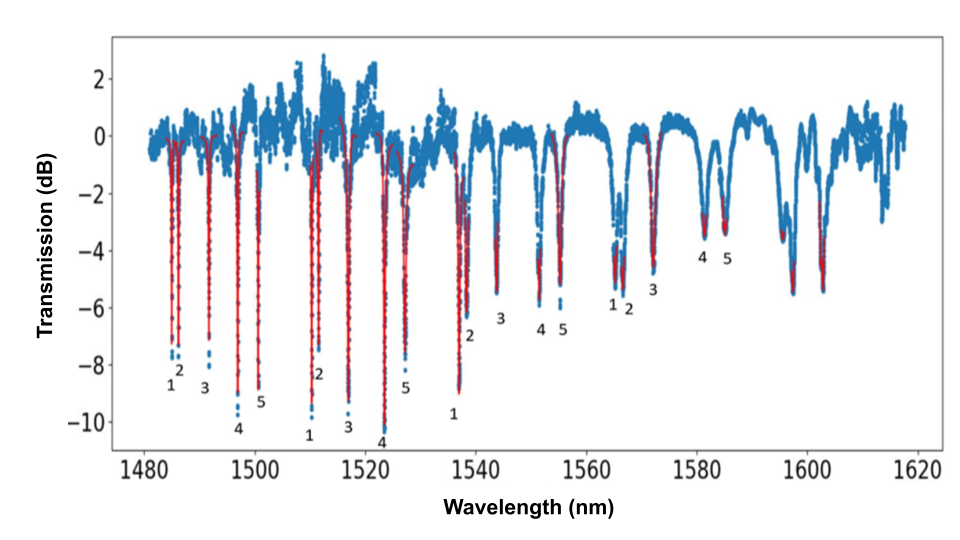}
\caption{Flattened spectrum taken of ANT-07 in the range of 1480 to 1620 nm. Each number below the dips indicates the ring resonator by which the dip was created. The blue curves are the spectrum that was transmitted to the InGaAs detector by ANT-07, while the red curves are dips that were constructed by a dip-fitting script. The order locations are approximately: first order (1480-1505 nm), second order (1510-1525 nm), third order (1537-1555 nm), \& fourth order (1565-1590 nm). The fifth order is indiscernible.}
\label{lowerH_spectra}
\end{figure}

The lower H-band spectrum contains the first four orders of dips produced by suppression in ANT-07 ($m=1,2,3,4$). Henceforth, the first four orders will be referred to as the lower H-band. The lower H-band spectrum produced by ANT-07 was flattened to account for broad features of the continuum. The spectrum was flattened by fitting a polynomial to segments of the continuum, and then flattening each individual segment relative to its local full illumination. This piecewise approach minimized the impacts of increasing attenuation with order. The flattened lower H-band spectrum is presented in Fig. \ref{lowerH_spectra} . 

\par 

The performance metrics used to characterize the dips created from each of ANT-07’s PRRs are: suppressed wavelength ($\lambda_{sup,m}$), full width at half maximum (FWHM), and depth. To extract the performance metrics from the spectrum, we created a script that found the lowest points of the spectrum and took them to be at the minima of intended dips. An inverse Lorentzian distribution was then fitted to these dips to identify their minima \citep{GuFang}. The minima of these Lorentzian fits were taken to be the suppressed wavelengths. The FWHM and depths for each dip were also extracted and recorded from the fit. The red curves in Fig. \ref{lowerH_spectra} indicate the inverse Lorentzian distributions that were predicted by the dip-fitting script, while the experimentally measured spectrum is in blue. Metrics predicted by this script were termed dip-fitted (DF) performance metrics. In addition to these DF metrics for ANT-07, we also recorded the FWHM and depth as determined by visual inspection. These visually-inspected (VI) metrics were recorded due to concerns about the accuracy of measurements made by the dip-fitting script, which is still the subject of ongoing development.
\par

Fig. \ref{lowerH_spectra} indicates that, as a dip's order increases, its FWHM also increases such that consecutive dips begin to blend together and become unresolvable. This is expected, as FWHM of a dip is proportional to the suppressed wavelength squared \citep{sellis_text}. The dips also become more shallow with increasing wavelength, likely a result of increased attenuation due to de-tuning \citep{microres_text}.  \par 

As seen in Fig. \ref{lowerH_spectra}, only the third dip avoids blending with its neighboring dips. The third dip’s lack of blending at higher orders allows for the most accurate extraction of performance metrics, both DF and VI. Accordingly, the performance metrics of the third dip are separated from those of the other dips. The values of both the DF and VI performance metrics for ANT-07's third dip in the lower H-band are tabulated in Table \ref{lowermetrics_dip3}. The performance metrics for the first, second, fourth, and fifth dips are tabulated in Table \ref{lowermetrics_all} of the Appendix. \par

\begin{deluxetable}{cccccc}
\label{lowermetrics_dip3}

%% Use \tablewidth{?pt} to over-ride the default table width.
%% This is the title of the table.
\tablecaption{DF and VI performance metrics of the third dip created by the third PRR in ANT-07 for the first four orders contained in the lower H-band. Refer to Table \ref{lowermetrics_all} in the Appendix for the tabulated performance metrics of the first, second, fourth, and fifth dips.}

%% This command over-rides LaTeX's natural table count
%% and replaces it with this number.  LaTeX will increment 
%% all other tables after this table based on this number
%\tablenum{1}
\tablehead{\colhead{m} & \colhead{DF} & \colhead{DF} & \colhead{VI} & \colhead{DF} & \colhead{VI} \\ 
\colhead{} & \colhead{$\lambda_{sup,m}$} & \colhead{depth} & \colhead{depth} & \colhead{FWHM} & \colhead{FWHM} \\ 
\colhead{} & \colhead{(nm)} & \colhead{(dB)} & \colhead{(dB)} & \colhead{(nm)} & \colhead{(nm)} } 
%% All data must appear between the \startdata and \enddata commands
\startdata
1 & 1491.696 & 7.121 & 8.0 & 0.1735 & 0.14 \\
2 & 1516.921 & 9.555 & 9.3 & 0.3450 & 0.30 \\
3 & 1543.763 & 2.140 & 5.4 & 0.3919 & 0.45 \\
4 & 1572.098 & 4.662 & 4.7 & 0.9552 & 0.85 \\
\enddata
\end{deluxetable}

The performance metrics considered going forward are the DF suppressed wavelength, VI depth, and VI FWHM. The VI depth and FWHM are selected in lieu of their DF counterparts because the DF script struggled to extract these metrics. Only the third dip has DF depths and FWHMs that are somewhat consistent with their VI counterparts (Table \ref{lowermetrics_dip3}). For all other dips, the DF depth and DF FWHM are inconsistent with their VI counterparts as early as the third order (Table \ref{lowermetrics_all}). This inconsistency indicates that the dip-fitting script failed to accurately extract depth and FWHM metrics, likely as a result of worsened blending with increasing order. Accordingly, polynomial fits for extrapolation must be constructed from the DF suppressed wavelength, VI depth, and VI FWHM of each dip in the lower H-band (Tables \ref{lowermetrics_dip3} and \ref{lowermetrics_all}). Particular emphasis is placed on the results from the third dip, as it is least impacted by blending. \par 

From each dip's selected metrics in the lower H-band (Tables \ref{lowermetrics_dip3} and \ref{lowermetrics_all}), polynomial fits are constructed using least-squares minimization to predict performance metrics at the sixth order ($m=6$) for all five dips. Henceforth, the sixth order is referred to as the upper H-band. For the DF suppressed wavelength ($\lambda_{sup,m}$) we attempt linear, quadratic-linear, quadratic-only, and cubic-quadratic-linear fits. For the extrapolation of the VI FWHM and VI depth, only linear fits are considered. A $\chi$-squared test is conducted to determine each model’s goodness of fit for predicting the performance metrics of all five dips through the lower H-band. The $\chi$-squared test statistic is given by Eq. \ref{chi_squared}.

\begin{equation}
\label{chi_squared}
\chi^2 = \sum \frac{(O_i - E_i)^2}{2\sigma_i^2}
\end{equation}

\noindent where $O_i$ is the observed metric, $E_i$ is the fitted metric, and $\sigma_i$ is the uncertainty on the observed metric. For the prediction of the DF suppressed wavelength, $\sigma$ is taken to be 10 pm based on the projected resolution of the laser and the granularity of the spectrum. For the prediction of the VI depth and the VI FWHM, $\sigma$ is taken to be 0.25 dB based on the ability to visually estimate the continuum level to then subtract the dip depth. All measurement uncertainties are assumed to be constant and not varying with order. The quadratic-only fit yielded a chi-squared value ($\chi^2$) on the order of  $10^{6}$, indicating the fit was highly inaccurate. The cubic-quadratic-linear fit yielded $\chi^2$'s on the order $10^{-20}$, indicating that the fit was highly over-constrained. Accordingly, these polynomial fits were excluded from analyses. The remaining $\chi^2$'s (Table \ref{lower_chisq}) indicate that the linear fit for the VI depth is possibly inaccurate and the linear fit for DF suppressed wavelength is almost certainly inaccurate. \par

\begin{deluxetable}{ccccc}
\label{lower_chisq}

%% Use \tablewidth{?pt} to over-ride the default table width.
%% This is the title of the table.
\tablecaption{$\chi^2$ values from computing the $\chi$-squared test statistic on the polynomial models constructed from each dip's performance metrics in the lower H-band (Tables \ref{lowermetrics_dip3} and \ref{lowermetrics_all}). All polynomial models were constructed by the least-squares minimization method.}

%% This command over-rides LaTeX's natural table count
%% and replaces it with this number.  LaTeX will increment 
%% all other tables after this table based on this number
%\tablenum{1}
\tablehead{\colhead{Dip \#} & \colhead{$\chi^2$ of } & \colhead{$\chi^2$ of} & \colhead{$\chi^2$ of } & \colhead{$\chi^2$ of } \\
\colhead{} & \colhead{ VI depth } & \colhead{VI FWHM} & \colhead{DF $\lambda_{sup,m}$} & \colhead{DF $\lambda_{sup,m}$} \\
\colhead{} & \colhead{ lin. fit } & \colhead{lin. fit} & \colhead{lin. fit} & \colhead{quad-lin. fit}} 
%% All data must appear between the \startdata and \enddata commands
\startdata
1 & 50.0 & 0.4 & 10,616 & 14.0 \\
2 & 2.0 & 0.4 & 10,817 & 7.5 \\ 
3 & 40.0 & 0.2 & 12,092 & 3.8 \\
4 & 40.0 & 0.04 & 14,184 & 37.0 \\
5 & 2.0 & 0.1 & 13,705 & 24.0 \\
\enddata
\end{deluxetable}

\begin{deluxetable}{cccc}
\label{wavelength_preds_exp}

%% Use \tablewidth{?pt} to over-ride the default table width.
%% This is the title of the table.
\tablecaption{Model-predicted suppressed wavelengths ($\lambda_{sup,m}$) of all dips at the sixth order compared with their experimental VI counterparts. The linear and quadratic-linear fits used for extrapolation were constructed from the DF suppressed wavelengths in Tables \ref{lowermetrics_dip3} and \ref{lowermetrics_all}.}

%% This command over-rides LaTeX's natural table count
%% and replaces it with this number.  LaTeX will increment 
%% all other tables after this table based on this number

\tablehead{\colhead{Dip} & \colhead{\hspace{1pt}Pred. $\lambda_{sup,m}$} & \colhead{Pred. $\lambda_{sup,m}$} & \colhead{VI $\lambda_{sup,m}$} \\
\colhead{\#} & \colhead{\hspace{1pt} lin. fit} & \colhead{\hspace{1pt}quad-lin. fit} & \colhead{\hspace{10pt}experimental} \\
\colhead{} & \colhead{\hspace{1pt}(nm)} & \colhead{\hspace{1pt}(nm)} & \colhead{\hspace{1pt}(nm)}}
%% All data must appear between the \startdata and \enddata commands
\startdata
1 & \hspace{1pt} 1617.96 & \hspace{1pt} 1625.96 & \hspace{1pt} 1625.50\\
2 & \hspace{1pt} 1619.50 & \hspace{1pt} 1627.59 & \hspace{1pt} 1627.50\\
3 & \hspace{1pt} 1624.93 & \hspace{1pt} 1633.48 & \hspace{1pt} 1634.00\\
4 & \hspace{1pt} 1636.82 & \hspace{1pt} 1646.07 & \hspace{1pt} 1648.00\\
5 & \hspace{1pt} 1640.56 & \hspace{1pt} 1649.66 & \hspace{1pt} 1651.25\\ 
\enddata
\end{deluxetable}

\begin{deluxetable}{ccc}
%% Use \tablewidth{?pt} to over-ride the default table width.
%% This is the title of the table.
\tablecaption{Model-predicted depths and FWHMs of all dips at the sixth order. The linear fits used for extrapolation were constructed from the VI depths and VI FWHMs presented in Tables \ref{lowermetrics_dip3} and \ref{lowermetrics_all}. Experimental VI depths and VI FWHMs were only successfully extracted for the third dip. Those values were 1 dB and 2.5 nm, respectively.}
\label{depth_fwhm_pred}
%\tablewidth{50pt}
%% This command over-rides LaTeX's natural table count
%% and replaces it with this number.  LaTeX will increment 
%% all other tables after this table based on this number
\tablehead{\colhead{Dip} & \colhead{\hspace{32pt}Pred. Depth} & \colhead{\hspace{32pt}Pred. FWHM} \\
\colhead{\#} & \colhead{\hspace{32pt}lin. fit} & \colhead{\hspace{32pt}lin. fit} \\
\colhead{} & \colhead{\hspace{32pt}(dB)} & \colhead{\hspace{32pt}(nm)}}
%% All data must appear between the \startdata and \enddata commands
\startdata
1 & \hspace{32pt} 4.54 & \hspace{32pt} 1.07 \\
2 & \hspace{32pt} 3.83 & \hspace{32pt} 1.31 \\
3 & \hspace{32pt} 2.02 & \hspace{32pt} 1.23 \\
4 & \hspace{32pt} -0.72 & \hspace{32pt} 1.40 \\
5 & \hspace{32pt} -0.39 & \hspace{32pt} 1.67 \\
\enddata
\end{deluxetable}

\begin{figure}[h!]
\centering
\vspace{3pt}
\includegraphics[scale=0.16]{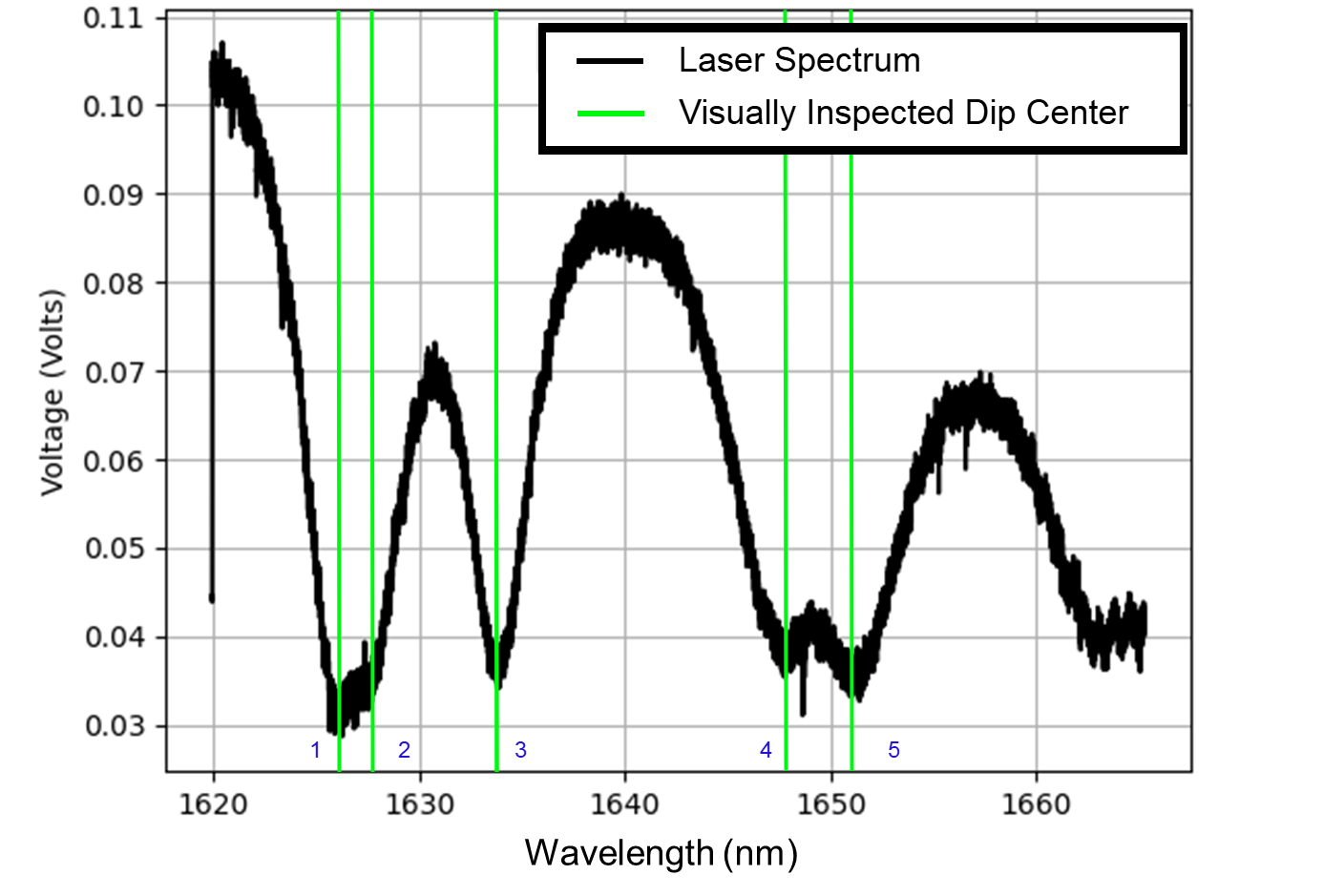}
\caption{Sixth order unflattened spectrum of ANT-07 taken in the range of 1620 nm to 1665 nm. Flattening was not achievable in the upper H-band due to dip broadening. Green lines indicate visually inspected dip centers ($\lambda_{sup,m}$) from Table \ref{wavelength_preds_exp} and are numbered according to their ring resonator in blue. Please note that, because the dip-fitter was not used to extract metrics, there is no inverse-Lorentzian fitted to the curves as had been done in Fig. \ref{lowerH_spectra}.} 
\label{upperH_spectra}
\end{figure}

We then extrapolate to predict the DF suppressed wavelength, VI depth, and VI FWHM for all dips at $m=6$ in the upper H-band (Tables \ref{wavelength_preds_exp} and \ref{depth_fwhm_pred}).  Note that the first, second, fourth, and fifth dips in the upper H-band spectrum (see Fig. \ref{upperH_spectra}) are broader and more difficult to resolve than they were for lower orders. As a result, there are insufficient regions of the spectrum at full illumination to construct segmented polynomial fits for the flattening of the spectrum. The unflattened spectrum is instead processed. Like the dips at $m=5$ in Fig. \ref{lowerH_spectra}, the DF script was not suited for this spectrum and did not yield meaningful results. \par

\begin{figure}[hbt!]
\centering
\vspace{3pt}
\includegraphics[scale=0.55]{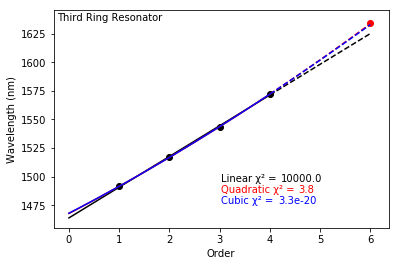}
\caption{The fits and extrapolations for the third dip with increasing order. Dashed lines indicate extrapolations for $m > 4$. Black is the linear fit, red is the quadratic-linear model, and blue is the cubic-quadratic-linear model. $\chi^2$s are tabulated in the bottom right. The red point at $m=6$ is the measured suppressed wavelength for ANT-07’s third dip, while black points are the measured suppressed wavelengths from the lower H-band.}
\label{thirddip_extrapolation}
\end{figure}

\begin{figure}[hbt!]
    \centering
    \includegraphics[width=.95\columnwidth]{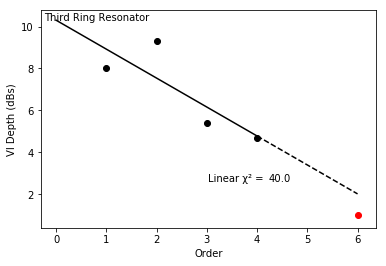}
    %\label{other_extrapolation:depth}
    \includegraphics[width=.95\columnwidth]{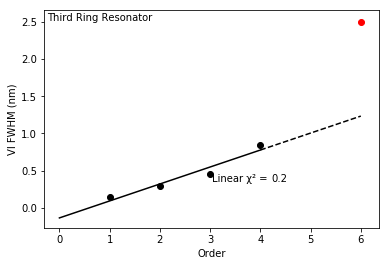}
    %\label{other_extrapolation:fwhm}
    \caption{Plots showing the linear fits and extrapolations for VI depth (top) and FWHM (bottom) of the third PRR’s dips with order. Dashed lines indicate extrapolations for $m > 4$. The red points at the 6th order are the experimental VI measurements for depth and FWHM, respectively.}
    \label{other_extrapolation:other_extrapolation}
\end{figure}

Thus, we determined performance metrics by visual inspection for the sixth order.  While the suppressed wavelength could be measured by visual inspection for all dips at the sixth order, the high degree of blending made it difficult to visually extract their depths and FWHMs. For this reason, VI depth and VI FWHM were measured for only the third dip due to its lack of blending. The third dip's VI depth and FWHM were measured as 1 dB and 2.5 nm, respectively. The predicted and experimental suppressed wavelengths for the sixth order are tabulated in Table \ref{wavelength_preds_exp}. Table \ref{depth_fwhm_pred} only contains the model-predicted depths and FWHMs. The predicted metrics for the third dip are compared with the experimental VI measurements in Figs. \ref{thirddip_extrapolation} \& \ref{other_extrapolation:other_extrapolation}.\par

\section{Discussion}
\label{sec:Discussion}

Based on these analyses, the best model for the prediction of suppressed wavelength with respect to order appears to be a linear-quadratic fit. This is indicated by Table \ref{lower_chisq}, for which the linear $\chi^2$ is far too large and the linear-quadratic-cubic $\chi^2$ is near zero. This is further corroborated by the experimental measurement of the suppressed wavelength for the third dip at the sixth order, seen in Fig. \ref{thirddip_extrapolation}. We suspect that the relationship between order and suppressed wavelength follows a quadratic curve as a result of the quadratic term that arises in the $FSR_{\lambda}$ according to Eq. \ref{free_spectral_range}. However, it should be noted that this equation for FSR assumes that the dispersion of light ($n_g$) is constant across all wavelengths. In practice, this is not true due to the material properties of the resonator \citep{sellis_text}. Based on work by a PhD student in the collaboration\footnote{P. Liu, private communication}, the group refractive index progresses in an approximately linear fashion with respect to the input wavelength of light for ANT-07. Accordingly, we shed the assumption that the group refractive index was constant by adding a linear wavelength-dependent dispersion term:

\begin{equation}
\label{lin_wvlngth_disp}
n_g(\lambda) = n_0 + n_1\lambda
\end{equation}

\noindent where $n_1$ is the linear wavelength dependent dispersion term, and $n_0$ is the base index of refraction. In this equation, $\lambda$ can be any wavelength of light injected into the ring resonator. However, only the resonant wavelengths ($\lambda_{res,m}$) are of practical interest. Eq. (\ref{free_spectral_range}) can then be substituted into Eq. (\ref{lin_wvlngth_disp}), and its Taylor series expansion yields: 

\begin{equation}
\label{taylor_series_expansion}
FSR_\lambda = \frac{\lambda_0^2}{Ln_0} \Biggl(1-1 \Bigl(\frac{n_1}{n_0}\lambda \Bigr) +\Bigl( \frac{n_1}{n_0}\lambda \Bigr)^2\Biggr)
\end{equation}

Because $\frac{n_1}{n_0}$ is not negligible, the result is that the FSR has linear, quadratic, and cubic terms that contribute to the location of the suppressed wavelength. There is a loss of accuracy between the fourth and fifth dips in our quadratic suppressed wavelength predictions, possibly indicating a need for a cubic term in the model. However, the $\chi^2$ from our attempted cubic fit indicates a high degree of over-fitting. The use of a cubic term for suppressed wavelength prediction requires further exploration.  \par 

The trends of the third dip for the VI depth and VI FWHM are much less certain. The $\chi^2$'s for depth vary quite drastically; goodness of fit varies from decent to poor. There is some evidence for a linear fit in the case of the third dip, as the measurement from the upper H-band spectrum seems to suggest a somewhat linear nature (see Fig. \ref{other_extrapolation:other_extrapolation}(top)). For FWHM, the $\chi^2$'s are sufficiently small, but the measured FWHM from $m=6$ does not corroborate this finding (see Fig. \ref{other_extrapolation:other_extrapolation}(bottom)). At present, it is clear that linear fitting alone cannot accurately predict the depth and FWHM for ANT-07 in the upper H-band. 

\section{Further Work}
\label{sec:further_work}
Currently, there is still much ongoing and future work before these PRR filters can be integrated into a telescope. Some immediate items include adjustments to the dip-fitting script and updates to the current experimental setup. 

\par Precision estimations of FWHM and depth will be necessary to achieve total line suppression in telescopes. Currently, the dip-fitting script's predictions of FWHM and depth do not align with VI measurements. The collaboration is presently working to improve the dip-fitting script by comparing the performances of different fitted distributions, such as the Voigt, as well as studying the impacts of spectrum flattening on the dip-fitter's extracted metrics. After improving the dip-fitters extraction of FWHM and depth, we will need to better characterize the relationships that FWHM and depth have with order.

\par There is also a need to improve the precision and reproducibility of polarization measurements. The current polarization tracking method does not yield a precise measure of polarization, but rather a discrete benchmark. Ongoing work in the group attempts to more finely sample the spectrum across angles of the in-line polarizer so that the dips' SNR can be maximized.

It is important to note that these ring resonator filters are still in an early stage of development and have not yet been integrated or tested on a telescope. Several long-term tasks remain before integration will be feasible. These include: 
\begin{enumerate}
    \item Narrowing ring resonators' dip depths to improve OH line suppression.
    \item Construction of a thermal controller to tune multiple dips' suppressed wavelengths on a chip.
    \item Design and construction of compact packaging for implementation in a telescope's optical path. 
\end{enumerate}

Upon successfully integrating these chips in a telescope, our group expects to either license the design to other observatories and individual astronomers, or to construct these systems for the astronomical 
community.

\par

\section{Conclusions}
\label{sec:conclusions}

In this paper, we characterized three performance metrics for a prototype photonic ring-resonator (PRR) filter device intended for the suppression of atmospheric OH emission lines. The PRR's output spectrum in the lower H-band ($m = 1 ,2,3,4$) was used to predict three performance metrics in the upper H-band ($m=6$): suppressed wavelength, depth, and full width at half maximum (FWHM).

To make these predictions, polynomial models were fitted from the lower H-band data and then used to extrapolate to the upper H-band. These models were evaluated from both a $\chi$-squared analysis and a comparison with experimentally determined performance metrics in the upper H-band. From this comparison, we find evidence that suppressed wavelengths progress with order according to a quadratic function. This relationship is consistent with the theoretical dependence of a PRR's free spectral range with increasing wavelength (and, therefore, order). This finding is a significant first step towards the prediction of a PRR's dip location and shape such that OH emission lines can be more accurately suppressed over a broader wavelength range. In contrast, no significant trends were identified for the progression of depth and FWHM with order. 
\par

Crucially, this paper demonstrates that PRR filters have the potential to be viable for sky background suppression in the near infrared. We successfully demonstrated that our current PRR filter performs as expected from the theory shown in Section \ref{sec:Discussion}. Additionally, we successfully quantified the coefficients in Eq.~\ref{free_spectral_range} and demonstrated the capabilities of our upper H-band testbed for PRR filters with a relatively straightforward analysis. Further, this work illuminates the challenges associated with utilizing higher order dips in a PRR filter for the suppression of sky background.
 
\section{Acknowledgments}
A. Hermann and R. Kehoe wish to thank SMU’s Undergraduate Research Assistanceship program and Department of Physics for support of this work. Computational resources for this research were provided by SMU’s O’Donnell Data Science and Research Computing Institute. Development of the ANT-07 device and initial measurements of its performance were supported by Argonne National Lab. The ANT-07 device was fabricated at Applied Nanotools, Inc., Canada. We wish to acknowledge Dave Underwood from Argonne National Lab for designing and fabricating the sensor in the InGaAs detector. Use of the Center for Nanoscale Materials, an Office of Science user facility, was supported by the U.S. Department of Energy, Office of Science, Office of Basic Energy Sciences, under Contract No. DE-AC02-06CH11357.
\par 

\newpage

\bibliography{references}

\begin{thebibliography}{}
\expandafter\ifx\csname natexlab\endcsname\relax\def\natexlab#1{#1}\fi
\providecommand{\url}[1]{\href{#1}{#1}}
\providecommand{\dodoi}[1]{doi:~\href{http://doi.org/#1}{\nolinkurl{#1}}}
\providecommand{\doeprint}[1]{\href{http://ascl.net/#1}{\nolinkurl{http://ascl.net/#1}}}
\providecommand{\doarXiv}[1]{\href{https://arxiv.org/abs/#1}{\nolinkurl{https://arxiv.org/abs/#1}}}

\bibitem[{{Avelino} {et~al.}(2019){Avelino}, {Friedman}, {Mandel}, {et~al.}}]{Avelino19}
{Avelino}, A., {Friedman}, A.~S., {Mandel}, K.~S., {et~al.} 2019, \apj, 887, 106, \dodoi{10.3847/1538-4357/ab2a16}

\bibitem[{{Barone-Nugent} {et~al.}(2012){Barone-Nugent}, {Lidman}, {Wyithe}, {et~al.}}]{barone-nugent_2012}
{Barone-Nugent}, R.~L., {Lidman}, C., {Wyithe}, J.~S.~B., {et~al.} 2012, \mnras, 425, 1007, \dodoi{10.1111/j.1365-2966.2012.21412.x}

\bibitem[{{Batista de Nazaré} {et~al.}(2017){Batista de Nazaré}, {da Silva Barros Allil}, \& {Werneck}}]{FBG_text}
{Batista de Nazaré}, F.~V., {da Silva Barros Allil}, R.~C., \& {Werneck}, M.~M. 2017, {Fiber Bragg Gratings: Theory, Fabrication, and Applications}, \dodoi{https://doi.org/10.1117/3.2286558.ch3}

\bibitem[{{Blanton} {et~al.}(2011){Blanton}, {Kazin}, {Muna}, {et~al.}}]{blanton_2011}
{Blanton}, M.~R., {Kazin}, E., {Muna}, D., {et~al.} 2011, \aj, 142, 31, \dodoi{10.1088/0004-6256/142/1/31}

\bibitem[{{Bogaerts} {et~al.}(2012){Bogaerts}, {De Heyn}, {Van Vaerenbergh}, {et~al.}}]{bogaerts_2011}
{Bogaerts}, W., {De Heyn}, P., {Van Vaerenbergh}, T., {et~al.} 2012, Laser \& Photonics Reviews, 6, 47, \dodoi{10.1002/lpor.201100017}

\bibitem[{{Brout} {et~al.}(2022){Brout}, {Scolnic}, {Popovic}, {et~al.}}]{brout_2022}
{Brout}, D., {Scolnic}, D., {Popovic}, B., {et~al.} 2022, \apj, 938, 110, \dodoi{10.3847/1538-4357/ac8e04}

\bibitem[{Chance {et~al.}(2005)Chance, Kurosu, \& Sioris}]{Chance_2005}
Chance, K., Kurosu, T.~P., \& Sioris, C.~E. 2005, Appl. Opt., 44, 1296, \dodoi{10.1364/AO.44.001296}

\bibitem[{{Ellis} {et~al.}(2023){Ellis}, {Bland-Hawthorn}, \& {Leon-Saval}}]{sellis_text}
{Ellis}, S., {Bland-Hawthorn}, J., \& {Leon-Saval}, S. 2023, {Principles of Astrophotonics}, \dodoi{10.1142/q0391}

\bibitem[{{Ellis} \& {Bland-Hawthorn}(2008)}]{scellis_2008}
{Ellis}, S.~C., \& {Bland-Hawthorn}, J. 2008, \mnras, 386, 47, \dodoi{10.1111/j.1365-2966.2008.13021.x}

\bibitem[{Fang {et~al.}(2020)Fang, Gu, Zheng, Zhao, Gan, \& Zhao}]{GuFang}
Fang, L., Gu, L., Zheng, J., {et~al.} 2020, Journal of Lightwave Technology, 38, 4429, \dodoi{10.1109/JLT.2020.2991648}

\bibitem[{Hecht(2002)}]{hecht}
Hecht, E. 2002, Optics, 4th edn. (Addison Wesley)

\bibitem[{{Heebner} {et~al.}(2007){Heebner}, {Grover}, \& {Ibrahim}}]{microres_text}
{Heebner}, J., {Grover}, R., \& {Ibrahim}, T. 2007, {Optical Microresonators: Theory, Fabrication, and Applications}, \dodoi{https://doi.org/10.1007/978-0-387-73068-4}

\bibitem[{Kuehn {et~al.}(2024)Kuehn, Kelley, Kuhlmann, Kehoe, Lasker, Rajeev, Hermann, \& Ellis}]{kkuehn_2024}
Kuehn, K., Kelley, T., Kuhlmann, S., {et~al.} 2024, in Advances in Optical and Mechanical Technologies for Telescopes and Instrumentation VI, ed. R.~Navarro \& R.~Jedamzik, Vol. 13100, International Society for Optics and Photonics (SPIE), 1310069, \dodoi{10.1117/12.3018485}

\bibitem[{{Kuehn} {et~al.}(2020){Kuehn}, {Kuhlmann}, {Ellis}, {et~al.}}]{kkuehn_2020}
{Kuehn}, K., {Kuhlmann}, S., {Ellis}, S., {et~al.} 2020, in Society of Photo-Optical Instrumentation Engineers (SPIE) Conference Series, Vol. 11451, Society of Photo-Optical Instrumentation Engineers (SPIE) Conference Series, 114516A, \dodoi{10.1117/12.2561990}

\bibitem[{{Liu} {et~al.}(2021){Liu}, {Czaplewski}, {Ellis}, {et~al.}}]{pliu_2021}
{Liu}, P., {Czaplewski}, D.~A., {Ellis}, S., {et~al.} 2021, \ao, 60, 3865, \dodoi{10.1364/AO.421383}

\bibitem[{{Morganson} {et~al.}(2018){Morganson}, {Gruendl}, {Menanteau}, \& {DES Collaboration}}]{morganson_2018}
{Morganson}, E., {Gruendl}, R.~A., {Menanteau}, F., \& {DES Collaboration}. 2018, \pasp, 130, 074501, \dodoi{10.1088/1538-3873/aab4ef}

\bibitem[{{Peterson, E. R.} {et~al.}(2024){Peterson, E. R.}, {Scolnic, D.}, {Jones, D. O.}, {et~al.}}]{Peterson24}
{Peterson, E. R.}, {Scolnic, D.}, {Jones, D. O.}, {et~al.} 2024, \aap, 690, A56, \dodoi{10.1051/0004-6361/202450052}

\bibitem[{Phillips(2012)}]{Phillips12}
Phillips, M.~M. 2012, Publications of the Astronomical Society of Australia, 29, 434–446, \dodoi{10.1071/AS11056}

\bibitem[{Robertson(2017)}]{Robertson_2017}
Robertson, J.~G. 2017, Publications of the Astronomical Society of Australia, 34, e035, \dodoi{10.1017/pasa.2017.29}

\bibitem[{{Rousselot} {et~al.}(2000){Rousselot}, {Lidman}, {Cuby}, {et~al.}}]{rousselot_2000}
{Rousselot}, P., {Lidman}, C., {Cuby}, J.~G., {et~al.} 2000, \aap, 354, 1134

\bibitem[{Wood-Vasey {et~al.}(2008)Wood-Vasey, Friedman, Bloom, {et~al.}}]{Wood-Vasey08}
Wood-Vasey, W.~M., Friedman, A.~S., Bloom, J.~S., {et~al.} 2008, The Astrophysical Journal, 689, 377, \dodoi{10.1086/592374}

\bibitem[{{Yoffe} {et~al.}(1995){Yoffe}, {Krug}, {Ouellette}, \& {Thorncraft}}]{yoffe_1995}
{Yoffe}, G.~W., {Krug}, P.~A., {Ouellette}, F., \& {Thorncraft}, D.~A. 1995, \ao, 34, 6859, \dodoi{10.1364/AO.34.006859}

\end{thebibliography}

%\section{Conflicts of Interest}
%The authors have no conflicts to disclose.

\newpage
\onecolumngrid
\appendix

%\centering \textbf{APPENDIX}

\begin{table}[h!]
\setlength{\tabcolsep}{0pt}
\label{appendix_table}

\centering
\caption{The performance metrics of the first, second, fourth, and fifth dips created by ANT-07 for the first four orders contained in the lower H-band. DF indicates the values that were determined by the dip-fitter, while VI indicates the values that were determined by visual inspection.}

\begin{ruledtabular}
\begin{tabular}{c c r r r r r}

\thead{Dip \\ \color{white}.} & \thead{Order\\ \color{white}.} & \thead{DF $\lambda_{sup,m}$ \\ (nm)} & \thead{DF depth \\ (dB)} & \thead{VI depth \\ (dB)} & \thead{DF FWHM \\ (nm)} & \thead{VI FWHM \\ (nm)} \\
\hline	% horizontal line to separate headings from data
1 & 1 & 1484.971 & 7.089 & 7.7 & 0.1282 & 0.12 \\
1 & 2 & 1510.302 & 8.419 & 9.8 & 0.1651 & 0.15 \\ 
1 & 3 & 1536.974 & 7.976 & 8.2 & 0.3236 & 0.35 \\
1 & 4 & 1565.218 & 1.068 & 5.2 & 0.4651 & 0.75 \\
2 & 1 & 1486.181 & 7.101 & 7.7 & 0.1230 & 0.11 \\
2 & 2 & 1511.555 & 7.247 & 7.5 & 0.1439 & 0.14 \\
2 & 3 & 1538.313 & 3.631 & 6.2 & 0.7279 & 0.51 \\
2 & 4 & 1566.629 & 0.958 & 5.4 & 0.3259 & 0.85 \\
4 & 1 & 1496.912 & 9.199 & 9.6 & 0.1692 & 0.14 \\
4 & 2 & 1523.454 & 9.970 & 10.2 & 0.2703 & 0.30 \\
4 & 3 & 1551.485 & 1.586 & 5.7 & 0.2066 & 0.45 \\
4 & 4 & 1581.392 & 0.752 & 3.5 & 0.6833 & 0.85 \\
5 & 1 & 1500.613 & 7.267 & 8.8 & 0.1130 & 0.14 \\
5 & 2 & 1527.174 & 6.778 & 7.6 & 0.4370 & 0.40 \\
5 & 3 & 1555.234 & 5.215 & 5.2 & 0.6946 & 0.65 \\
5 & 4 & 1585\footnote{For the fourth order of the fifth dip, the fitter failed to find the dip. The VI suppressed wavelength has been included in lieu of the DF suppressed wavelength.} & $\cdot\cdot\cdot$ & 3.3 & $\cdot\cdot\cdot$& 1.10 \\

\end{tabular}
\end{ruledtabular}
\label{lowermetrics_all}
\end{table}

\end{singlespace}
\end{document}